\documentclass[prd,letterpaper,superscriptaddress,twocolumn]{revtex4}%
\usepackage{graphicx}
\usepackage{bm}
\usepackage{latexsym}
\usepackage{epsf}
\usepackage{rotating}
\usepackage{epsfig,graphics,rotate,color}
\usepackage{wrapfig}
\usepackage{amssymb}
\usepackage{amsmath}
\usepackage{amsfonts}
\usepackage{subfigure}
\usepackage{array,hhline,dcolumn}%
\usepackage[normalem]{ulem}
\providecommand{\U}[1]{\protect\rule{.1in}{.1in}}
\begin{document}
\title{Coherent Neutrino Scattering in Dark Matter Detectors}
\author{A.J. Anderson}
\affiliation{Massachusetts Institute of Technology, Cambridge, MA 02139, USA}
\author{J.M. Conrad}
\affiliation{Massachusetts Institute of Technology, Cambridge, MA 02139, USA}
\author{E. Figueroa-Feliciano}
\affiliation{Massachusetts Institute of Technology, Cambridge, MA 02139, USA}
\author{K. Scholberg}
\affiliation{Duke University, Durham, NC 27708, USA}
\author{J. Spitz }
\affiliation{Yale University, New Haven, CT 06520, USA}

\begin{abstract} 

%  Although never observed, coherent neutrino- and WIMP-nucleus interaction signatures are expected to be quite similar. This
%  paper describes how a next generation ton-scale dark matter detector
%  could discover neutrino-nucleus coherent scattering, a precisely predicted Standard Model process, and  use the events to enhance their WIMP search and analysis.  Recently proposed for oscillation physics at
%  underground laboratories, a high intensity pion- and muon- decay-at-rest neutrino beam would provide the source for these measurements.  In this paper, we calculate raw-rates for
%  various target materials commonly used in dark matter detectors and 
%  show that discovery of this interaction is possible in less than
%  xxx years. Furthermore, we discuss the possibility of coherent-neutrino-based detector calibration and physics possibilities with a GEODM module placed within tens of meters of the neutrino source.

Coherent elastic neutrino- and WIMP-nucleus interaction signatures are expected to be quite similar. This paper discusses how a next generation ton-scale dark matter detector could discover neutrino-nucleus coherent scattering, a precisely-predicted Standard Model process. A high intensity pion- and muon- decay-at-rest neutrino source recently proposed for oscillation physics at underground laboratories would provide the neutrinos for these measurements. In this paper, we calculate raw rates for various target materials commonly used in dark matter detectors and show that discovery of this interaction is possible with a 2~ton$\cdot$year GEODM exposure in an optimistic energy threshold and efficiency scenario. We also study the effects of the neutrino source on WIMP sensitivity and discuss the modulated neutrino signal as a sensitivity/consistency check between different dark matter experiments at DUSEL. Furthermore, we consider the possibility of coherent neutrino physics with a GEODM module placed within tens of meters of the neutrino source.

\end{abstract}
\maketitle

\section{Introduction}

Coherent elastic neutrino-nucleus scattering is a well-predicted Standard
Model interaction that is of wide-spread interest. Characterizing the process can provide a probe of beyond the Standard Model physics, such as through
non-standard neutrino interactions~\cite{kate}, and can contribute to
our understanding of supernova dynamics. Although
experiments have been proposed to search for this channel~\cite{CLEAR, CLEAR2, Cogent, texono}, neutrino-nucleus coherent elastic
scattering has never been observed.
This paper shows that, for existing proposed layouts of neutrino sources
and ton-scale dark matter detectors, discovery of coherent neutrino
scattering can occur with a 2~ton$\cdot$year exposure under an optimistic detection scenario. Along with coherent neutrino physics, the observation of rare, WIMP-like events with a well predicted flux (shape and absolute normalization) and cross section in a well-known time window can provide a cross-check of the sensitivity and efficiency of dark matter detectors. Furthermore, in the case that a substantial event sample is collected with a dedicated experiment close to the neutrino source, sensitivity to physics beyond the Standard Model and a unique probe of $\sin^{2}\theta_{W}$ can be obtained through a coherent neutrino scattering cross section measurement.

We discuss the physics importance of the coherent neutrino interaction in the next section and detection methods in Section~\ref{sec:detection}. Section~\ref{sec:discovery} provides our assumptions for a few example experiments and the raw rates in these detectors with various exposures and as a function
of envisioned energy thresholds and baseline lengths. The effect of coherent events as a background for WIMP interactions is also considered. Then, we consider the scenario in which a detector module is brought within tens of meters of the neutrino source and used to obtain a high statistics sample of events for %true dark matter detector calibration purposes (Section~\ref{sec:closedetector1}) and 
coherent neutrino physics (Section~\ref{sec:closedetector2}).

\section{Coherent neutrino scattering}
The coherent scattering cross section, $\sigma$, depends on the number of neutrons, $N$,
and protons, $Z$, of the target material with mass $M$.  If $T$ is the
recoil energy of the interaction and the incoming neutrino has energy $E_\nu$,
\begin{equation}
{{d\sigma}\over{dT}} = {{G_F^2}\over {4\pi}} Q_W^2 M \left(1-
{{MT}\over{2E_\nu^2}}\right) F(Q^2)^2. \label{coherent}
\end{equation}

In this equation, $G_F$ and $Q_W$ are the precisely known Fermi constant and weak
charge [$Q_W=N-(1-4~\mathrm{sin}^{2}\theta_{W})Z$], respectively.  The form factor, $F(Q^2)$, dominates the $\sim$5\% cross section uncertainty~\cite{Horowitz:2003cz}. 

As the cross section is well predicted, coherent elastic neutrino-nucleus scattering is an ideal source
to search for new physics in the neutrino sector.  A cross section
measurement with $\sim$10\% uncertainty will result in an uncertainty on
$\sin^{2}\theta_{W}$ of $\sim$5\%~\cite{kate}.  While this uncertainty
is large compared to existing and planned precision atomic parity violation and M\o ller scattering measurements, a discrepancy from the Standard Model prediction already
observed in the neutrino sector by the NuTeV experiment~\cite{NuTeV} motivates more neutrino-based measurements. Notably, a $\sin^{2}\theta_{W}$ measurement with coherent neutrino-nucleus scattering would be at $Q\sim\mathrm{0.04}$~GeV/c, well away from all previous neutrino scattering measurements (including NuTeV's at $Q\sim\mathrm{4}$~GeV/c).

A coherent neutrino-nucleus scattering cross section measurement agreement
within 10\% uncertainty of the Standard Model prediction will
result in limits on non-standard neutrino interactions (NSI) which improve on the present
ones by more than an order of magnitude~\cite{Barranco:2005yy,Barranco:2007tz,kate}. The low-$Q$ existing and planned precision measurements mentioned above are not sensitive to new physics unique to neutrino interactions. NSI terms can enter the Standard Model Lagrangian through an extra term, 
\begin{equation}
{\cal L}_\mathit{eff}^\mathit{NSI} =- \varepsilon^{fP}_{\alpha \beta} 2 \sqrt{2} G_F
 (\bar{\nu}_\alpha \gamma_{\rho} L \nu_\beta) (\bar{f} \gamma^{\rho}P
 f)
\end{equation}
where $f$ is a first generation Standard Model fermion, $e$, $u$, or $d$ and $P = L~\mathrm{or}~ R$~\cite{davidson}. These $\varepsilon^{fP}_{\alpha \beta}$ terms can appear due to a range of sources, including incorporating neutrino mass into the Standard Model~\cite{Schecter} and supersymmetry~\cite{Barger}. The $\varepsilon_{ee}$ and $\varepsilon_{e\tau}$ terms, in particular, are poorly constrained by existing measurements. Measuring a coherent scattering cross section in disagreement with the Standard Model expectation could be an indication of NSI. In the case that a cross section discrepancy is observed, multiple nuclear targets could be employed in order to disentangle effects from NSI, a $\sin^{2}\theta_{W}$ anomaly (e.g. consistent with NuTeV), and/or nuclear physics. 

Characterizing neutrino coherent scattering 
is also essential to the understanding of supernova evolution as the energy carried away by neutrinos comprises $\sim$99\% of the supernova's total energy and the coherent channel's cross section exceeds all others by at least an order of magnitude in the relevant energy region. 
In a stellar core collapse, the density of the electron/nucleus plasma at the core can reach $>$$10^{12}$~$\mathrm{g/cm^{3}}$. At these densities, a 20~MeV neutrino's mean free path is on the order of 0.5~km~\cite{Horowitz:1996ci} with the opacity in nucleus-rich regions dominated by coherent scattering.
% The process is especially important during infall when most of the nucleons are %still inside nuclei~\cite{bethe}. In such a neutrino-opaque environment and with %a cross section proportional to $\mathrm{A}^{2}$, coherent interactions may %push heavier elements to the outside of the shell and drastically affect the %explosion~\cite{freedman}. 
Along with being relevant for supernova evolution in general, the coherent cross section may also affect the supernova neutrino signals expected on Earth. 

The coherent process is important for supernova burst neutrino detection as well, providing information about all flavors of neutrinos--not just $\nu_{e}$/$\overline{\nu}_{e}$. For coherent neutrino-nucleus scattering specifically, unlike other channels of flavor-blind neutral current interactions, the nuclear recoil energy is proportional to neutrino energy due to the elastic nature of the interaction. Such information could be combined with charged current $\nu_{e}$/$\overline{\nu}_{e}$ interaction information from other sources to develop a complete picture of oscillations with supernova neutrinos. Note that approximately seven
neutrino-nucleus coherent events in one ton of Ar during a ten second
window for a galactic core-collapse supernova at 10~kpc are expected with a recoil energy threshold of 5~keV~\cite{Horowitz:2003cz}. Although the detectors discussed below are probably too small to provide a sizable sample of supernova burst neutrino-nucleus coherent scatters (unless the supernova is very close), an accelerator-based measurement to confirm the predicted interaction cross section would prove valuable to a next-generation coherent neutrino scattering experiment's supernova burst neutrino measurement.

\section{Detection}
\label{sec:detection}
The coherent neutrino-nucleus cross section favors very low recoil energies, in
the few-to-tens of keV range. This is well below the threshold of the most
sensitive recent and existing large-scale low energy neutrino detectors, like SNO~\cite{SNO}, Borexino~\cite{borexino}, and KamLAND~\cite{Kamland}, which explains why this relatively high cross section process has not yet
been observed. 
Dark matter detectors, on the other-hand, have energy thresholds in the $\sim$10~keV range and lower. As such, these detectors are potentially ideal coherent neutrino scattering detectors if given
a sufficiently large target mass and neutrino flux. A ton-scale dark matter detector at its nominal depth underground of ~1-2 km, in combination with an intense decay-at-rest (DAR) neutrino source, could discover the coherent neutrino scattering process. 

It is worth noting that dark matter
detectors could observe $^8$B solar
neutrino coherent events, as has been pointed out in Ref.~\cite{jocelyn}. The $^8$B rate depends on the detector threshold and material, with more events for lighter target nuclei and lower thresholds.
% with a ton$\cdot$year exposure and a 5~keV recoil energy threshold~\cite{jocelyn}. This can be compared to the 5-25 dark matter events/ton/year expected for a WIMP with a mass of 100~GeV/cm$^{2}$ and a cross section of $10^{-46}~\mathrm{cm}^{2}$.  
%Depending on the actual WIMP cross section, the detector material, and the energy threshold, solar neutrinos could become an irreducible background in dark matter detectors unable to reconstruct the initial WIMP's direction~\cite{jocelyn}. 
This solar neutrino signal becomes negligible for high-$A$ targets when the low-energy recoil threshold of the experiment is between 5 and 10~keV or more. There are zero solar coherent events expected with a 5/10~keV threshold for the targets (Xe, Ge/Ar) and exposures considered here. In contrast, the DAR accelerator source described in Section~\ref{sec:discovery} will produce a significant number of events within a well-defined time window when the accelerator is on.
These events can therefore only be considered a WIMP background during those times (see Sec.~\ref{sec:WIMPsearch}). In terms of a coherent neutrino physics measurement, the accelerator-based case has the luxury of an \textit{in-situ} background measurement (when the source is off) in addition to the higher-energy, above-threshold nuclear recoils. 

%A sample of xxx positively identified coherent neutrino scatters would make an ideal calibration source with which to test and tune
%dark matter detectors. Such a sample could be acquired with a dark matter detector or detector module placed within tens of meters of a surface accelerator.  Calibration of dark matter detectors is 
%important, difficult, etc...
%{\bf paragraph by Tali on general concerns for dark matter 
%detectors, why calibration is important, and why neutrons 
%are not ideal compared to coherent neutrino scatters}. 

\section{Rates and Time to Discovery}
\label{sec:discovery}
In order to be reasonably concrete, we study a set of experimental
designs inspired by proposals for the Deep Underground Science and
Engineering Laboratory (DUSEL).  We note that the detector designs are
not very different from those under consideration at other underground
laboratories and that the results can be easily scaled.  For the
neutrino source, we assume a DAR configuration produced by high
intensity cyclotrons which are now under development~\cite{Luciano}\cite{Jose} and proposed for DUSEL~\cite{EOI}.

\begin{figure}[b]\begin{center}
{\includegraphics[width=3.5in]{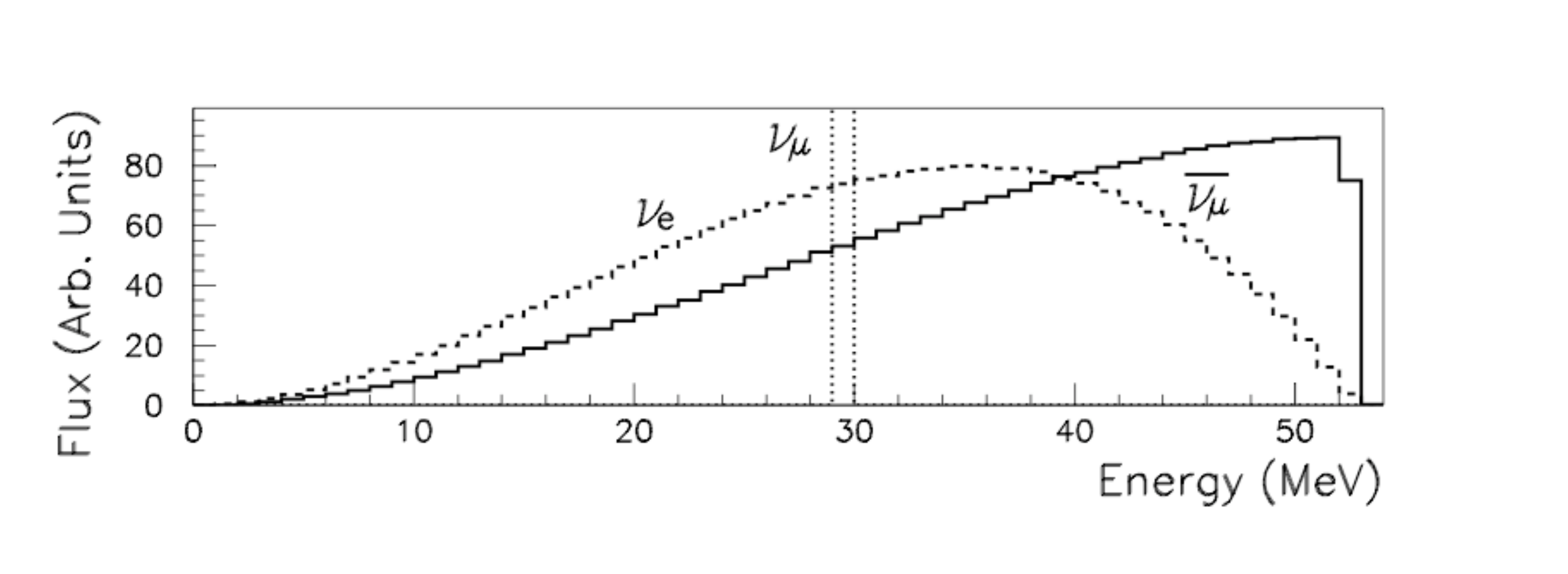}
} \end{center}
\vspace{-.8cm}
\caption{Energy distribution of neutrinos in a DAR source, from Ref~\cite{multi}.
\label{flux} }
\end{figure}

DAR neutrinos are known to be an excellent source for neutrino-nucleus coherent scattering experiments~\cite{CLEAR, CLEAR2}.
The neutrinos are produced with relatively low energies ($<$52.8 MeV), a range 
where coherent neutrino scattering dominates all
other cross sections by about an order of magnitude. Ref.~\cite{EOI} calls for a design with 800~MeV protons, accelerated via high intensity
cyclotron(s), which impinge on a carbon target and facilitate the production and eventual decay of pions: $\pi^{+}\rightarrow\mu^{+}\nu_{\mu}$, followed by $\mu^+
\rightarrow e^{+}\bar{\nu}_{\mu}\nu_{e}$.  A DAR source flux profile is shown in Fig.~\ref{flux}. 

High intensity DAR sources are being proposed for $CP$ violation
searches involving Gd-doped ultra-large water Cerenkov detectors at
underground science laboratories~\cite{multi, beamconfig, Argawalla1}. The design for this search~\cite{EOI} calls for multiple accelerator sites at varying distances from ultra-large water Cerenkov detector(s). The DAE$\delta$ALUS cyclotron-based near accelerator is proposed to run with a duty factor between 13\%~\cite{Luciano} and 20\%~\cite{EOI}, with an average of 1 MW of power in either case. Each 1 MW accelerator will provide $4 \times 10^{22}$ neutrinos of each flavor per year
produced as an isotropic flux within the time window, expected to be about 67~ms out of 500~ms for the near accelerator with the current design~\cite{janet}. The absolute normalization of the neutrino flux, determined with electron-neutrino elastic scattering ($\nu_{e}e^{-} \rightarrow \nu_{e}e^{-}$) as measured by ultra-large water Cerenkov detector(s), will have a systematic uncertainty of 1\% with dominant contributions from the cross section and energy scale uncertainties~\cite{multi}. The statistical uncertainty on the flux depends on the run period, but is expected to be on the order of 1\% as well. The near accelerator site is envisioned at or near the surface of the laboratory with the other cyclotrons
located many kilometers away. Note that the far sites
will produce insignificant coherent rates due to the 1/$\mathrm{r}^{2}$ dependence of the flux.  However, the near accelerator can provide a
significant event rate, during the 13\% beam-on time, for detectors
which are sufficiently close and large.  Examples of other
physics opportunities with this near accelerator are discussed in~Refs.~\cite{EOI, Argawalla2, Argawalla3}.

In order to provide realistic calculations, we examine three dark
matter experiments which are drawn from the designs of GEODM~\cite{GEODM}, 
LZ~\cite{LZ}, and MAX~\cite{MAX}.  These experiments use
germanium, xenon, and argon as their targets, respectively. Note that neon is also commonly considered as an alternative target medium in the noble liquid detectors mentioned~\cite{CLEAN}. We assume that the accelerator and beam dump are located at or near the surface. As GEODM is proposed for the DUSEL 7400~ft level and LZ/MAX are proposed for the 4800~ft level, we simply consider baseline lengths of 2.3~km and 1.5~km, respectively. The rates for each target are calculated for a ton$\cdot$year fiducial exposure since the design of each detector is still under consideration.

\begin{figure}[t]\begin{center}
\vspace{-.5cm}
\includegraphics[width=2.9in,angle=90]{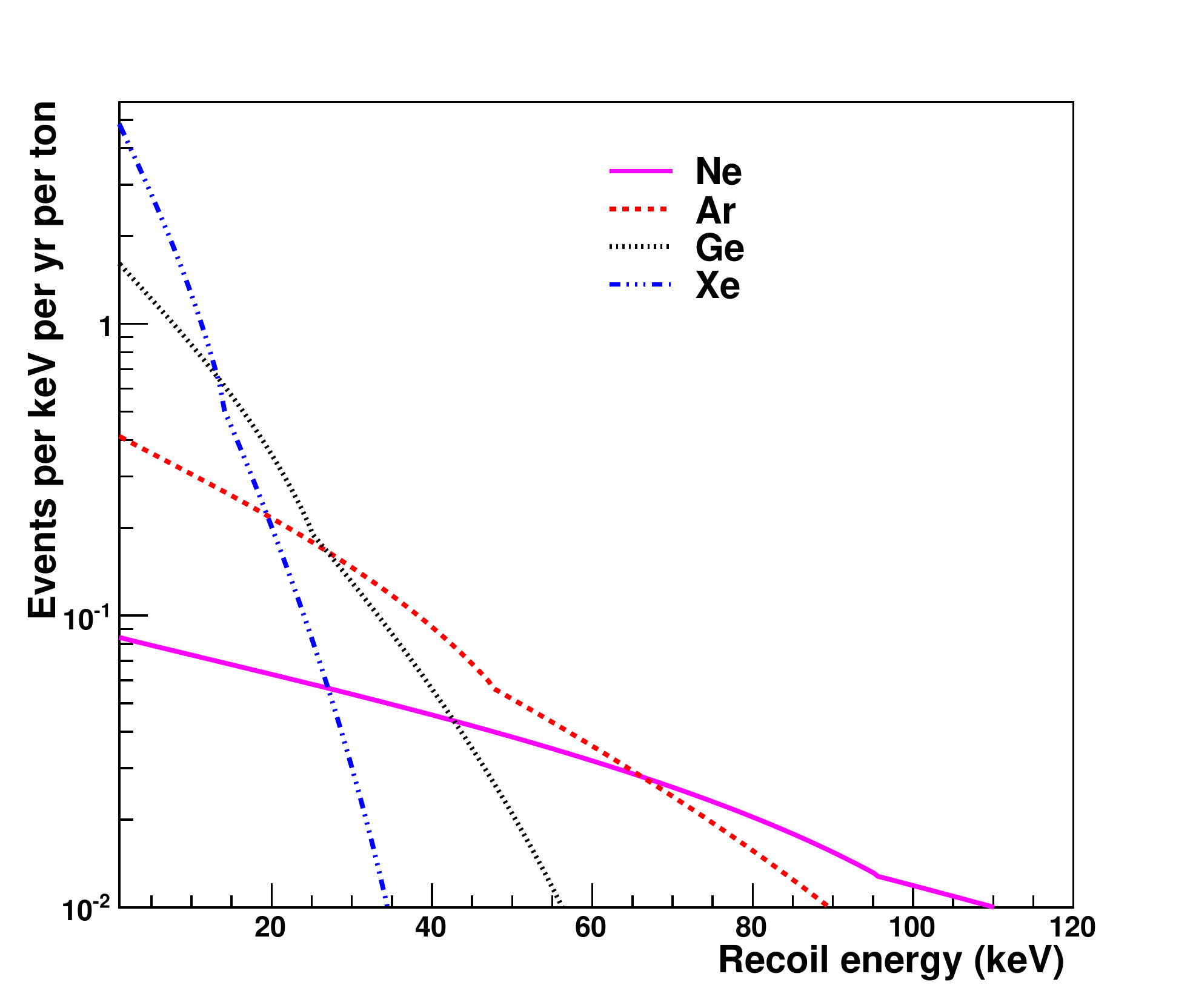}
\end{center}
\vspace{-.5cm}
\caption{Recoil energy distributions for coherent scattering 1.5~km from a DAR neutrino source for Ne, Ar, Ge, and Xe. The rates reported assume 100\% detection efficiency.}\label{recoils}
\end{figure}

As discussed above, the coherent neutrino-nucleus
interaction takes place at very low recoil energies.    Fig.~\ref{recoils}
shows the distribution of recoil energies for a DAR source with 
$^{20}$Ne, $^{40}$Ar, $^{76}$Ge,  and $^{132}$Xe. 
The experimental rates will strongly depend upon the recoil
energy threshold for reconstructed events, $T_{min}$. As the exact values
of this cut for the various detectors are unknown, we
consider five possible values of $T_{min}$. The rates with each cut are
obtained by summing Eq.~\ref{coherent} in bins of recoil energy $T$,
starting at $T_{min}$. For the aforementioned targets, we find the rates given in
Table~I, where we assume 100\% efficiency for detecting events in
the time-window above the threshold $T_{min}$.

\begin{table}[t]
\label{ratesa}
  \begin{center}
    {\footnotesize
      \begin{tabular}{|c|c|c|c|c|c|c|} \hline 
        Events/ton/year &  For  & \multicolumn{5}{c|}{$T_{min}$} \\ 
        at distance & target   & 0 keV & $5$ keV & $10$ keV & $20$ keV & $30$ keV \\ \hline
        1.5 km & $^{40}$Ar  & 11.1 & 9.1 & 7.5 & 4.9 & 3.1  \\
        &$^{132}$Xe & 36.4 & 16.3 & 6.6 & 1.1 & 0.1  \\ 
        &$^{76}$Ge$^{\dagger}$ & 21.9 & 14.6 & 9.4 & 3.5 & 1.4  \\ \hline
        2.3 km & $^{76}$Ge  & 9.3 & 6.2 & 4.0 & 1.5 & 0.6  \\ \hline %\hline
%        Solar & $^{40}$Ar  & 805 & 21.0 & $<$1.0 & - & -  \\
%        &$^{132}$Xe & 2617 & $<$1.0 & $<$1.0 & - & -  \\
%        & $^{76}$Ge  & 1495 & $<$1.0 & $<$1.0 & - & -  \\ \hline
      \end{tabular} 
      \caption{Coherent neutrino scattering events/ton/year (with the accelerator running at 1~MW with a 13\% duty factor) for various detector layouts and thresholds. The rates reported assume 100\% detection efficiency. $^{\dagger}$The present plan is for the GEODM ($^{76}$Ge-based) baseline to be 2.3~km--although 1.5~km is included for completeness.}
}

\end{center}

\end{table}

We note that the coherent event rates for dark matter detectors at their nominal depths underground are in the 0-35~events/ton/year range depending on target, baseline, energy threshold, and unrealistically assuming 100\% detection efficiency.  This is too low to be competitive with presently used neutron sources for detector calibration. Neutrons are adequate for energy calibration, despite their propensity to multiple scatter and activate the detector. Observing excess nuclear recoil events between beam-on and beam-off times would allow a dark matter search to cross-check its sensitivity. Unlike neutrons, neutrinos have a negligible probability of multiply scattering and neutrino interactions would be uniformly distributed throughout the detector; furthermore, unlike a neutron source, the measurement is noninvasive. A coherent neutrino signal would enable the study of the systematics in the expected scaling of the event rate across multiple dark matter experimental targets at the site.

\subsection{Detection at a Ton-Scale Dark Matter Experiment}
Next, we consider one of the nuclear targets mentioned above ($^{76}$Ge) in more detail. GEODM is a proposed ton-scale dark matter detector~\cite{GEODM} based on the cryogenic Ge crystal technology used in the CDMS experiment~\cite{CDMS}. The target design for GEODM is an array of 300 $\sim$5~kg Ge crystals operated at 40~mK with a total target mass of $\sim$1500~kg. Interaction events in an individual crystal produce populations of athermal phonons and electron-hole pairs which are measured by various phonon and ionization sensors lithographically patterned on the crystal surfaces. The ratio of ionization to phonon signals for an event is a powerful discriminator between electron and nuclear recoils.  The signals also enable precise determination of the position and energy of each event, which allow volume and energy cuts. This information is used to set the number of electron recoils that can pass the cuts and pose as nuclear recoils.  This electron recoil ``leakage'' into the nuclear recoil band constitutes one source of background events. There is also a background from muon- and radiogenic-induced neutrons, which is controlled through the use of radio-pure materials and passive and active shields to be $< 0.15$~events/ton/year.  

\begin{figure}[t]\begin{center}
\includegraphics[width=3.2in]{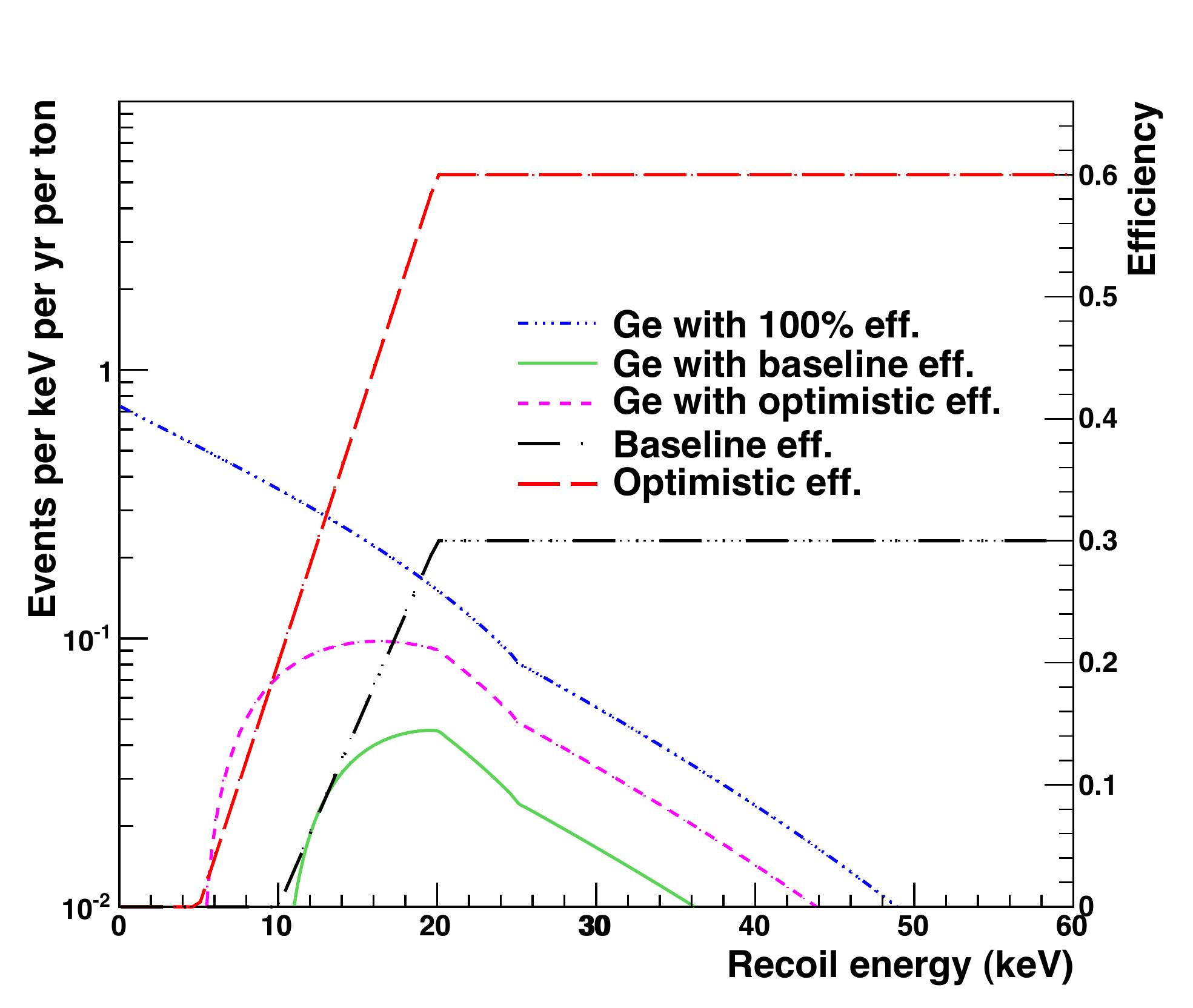}
\end{center}
\vspace{-.7cm}
\caption{Recoil energy distribution for coherent neutrino scattering on Ge at 2.3~km from a DAR neutrino source, before and after the effect of two hypothetical efficiency curves.}\label{efficiency}
\end{figure}

The efficiency of GEODM depends critically on the cuts used to achieve a target leakage background.  This is largely a function of the future detector's performance, which we need to estimate.  In light of this, we consider a ``baseline" scenario and an ``optimistic" scenario for the detector parameters, summarized in Table~II.  Fig.~\ref{efficiency} shows the recoil energy distribution in a detector with two hypothetical efficiency curves as a function of recoil energy.  The baseline scenario has a 10~keV nuclear recoil energy threshold with the efficiency rising linearly to 0.3 at 20~keV.  We additionally assume an energy resolution near threshold of 300~eV and a fiducial mass uncertainty of 5\%.  These parameters are consistent with the performance of the Ge detectors used in the CDMS II experiment~\cite{CDMSIIEff}.  The optimistic scenario assumes that refinement of the detectors improves all of these parameters.  In the optimistic scenario, we assume a 5~keV nuclear recoil energy threshold and an efficiency that rises linearly to 0.6 at 20~keV.  We also assume an energy resolution near threshold of 50~eV and a fiducial mass uncertainty of 1\%. The optimistic scenario parameters are anticipated to be achieved by detectors in the SuperCDMS experiment. In all cases, unless otherwise specified, we assume a raw exposure of 4.5 ton$\cdot$year before the efficiency curve is applied.  We also assume a leakage background of 1 event per 4.5 ton$\cdot$year raw exposure with a spectrum of $e^{-E_\textrm{recoil} / 10\textrm{ keV}}$, and a neutron background of 0.15~events per ton$\cdot$year of exposure with a spectrum of $e^{-E_\textrm{recoil} / 50\textrm{ keV}}$ before the efficiency curve is applied.  Events are required to have recoil energies less than 100~keV to be considered nuclear recoils, so we do not use any background events with energies exceeding 100~keV.  This neutron spectrum is convolved with each efficiency curve and scaled by the exposure to obtain the expected distribution of neutron events for each scenario.  For the baseline scenario we expect 0.13 total neutron events, and for the optimistic scenario we expect 0.29 total neutron events with a 4.5 ton$\cdot$year exposure.  

\begin{table}[t]
\label{base_opti}
  \begin{center}
    {\footnotesize
      \begin{tabular}{|c|c|c|} \hline 
        &Baseline & Optimistic \\ \hline
        Threshold & 10~keV & 5~keV \\
        Efficiency & See Fig. \ref{efficiency}   & See Fig. \ref{efficiency} \\ 
        Energy resolution near threshold&300~eV& 50~eV\\
        Neutron background events/(4.5~ton$\cdot$year)&0.13&0.29 \\
        Surface background events/(4.5~ton$\cdot$year)&1.0 &1.0 \\
        Fiducial mass uncertainty&5\%&1\%  \\ \hline

      \end{tabular} 
      \caption{The GEODM detector scenarios considered. The neutron background expectation is after efficiency corrections.}\label{GEODMscenarios}
}
\end{center}

\end{table}

Using the baseline efficiency and energy threshold, we find that a coherent neutrino rate of 0.8~events/ton/year is expected in a $^{76}$Ge-based detector at a baseline of 2.3~km. GEODM expects an overall leakage rate of about 1~background surface leakage event with a 4.5~ton$\cdot$year exposure, obtained by adjusting efficiency/fiducial volume cuts to get to that number. In addition, there is a neutron-induced nuclear recoil background (absolute rate convolved with detection efficiency) of 0.13 events/(4.5~ton$\cdot$year) and 0.29 events/(4.5~ton$\cdot$year) for the baseline and optimistic scenarios, respectively. The total expected background rate during the 13\% of beam-on time would therefore be 0.15~events/(4.5~ton$\cdot$year) for the baseline and 0.17~events/(4.5~ton$\cdot$year) for the optimistic scenario. The assumptions for GEODM as a coherent neutrino detector are shown in Table~III. Based on these considerations and assuming no WIMP ``background", one can see that an experiment like GEODM could find evidence for coherent scattering in a $\sim$4.5~ton$\cdot$year exposure with 3-4~signal events above a background expectation of 0.15~events. The probability for 4~observed events to be completely due to background, with a background expectation of 0.15 events, is $\sim$0.002\%. The signal rate and evidence/discovery timeline would be quickly improved in the case that the baseline efficiency estimate, especially in the low energy region (and possibly below 10~keV), is too conservative. With the optimistic energy threshold and efficiency scenario, we expect a coherent rate of 2.0~events/ton/year. Under the same background assumption as above, we find that GEODM will discover coherent scattering in a $\sim$2~ton$\cdot$year exposure with 4~signal events above a background expectation of 0.07 events. The probability for 4~observed events to be completely due to background, with a background expectation of 0.07 events, is $\sim$0.00009\%. 

It is worth noting that the absolute (100\% efficiency and 100\% on-time) solar coherent neutrino interaction rate on a $^{76}$Ge target above 5~keV is expected to be 0.079~events/ton/year~\cite{gutlein}. In either efficiency scenario and with 13\% beam-on time, the solar coherent ``background" rate is negligible.

%With 100\% efficiency and a 10~keV energy threshold, a 2.3~km GEODM experiment will collect a coherent sample approaching 40~events in about ten years. Statistical uncertainty will dominate the cross section measurement with such a sample as the systematic uncertainty can be assumed at the few percent level, in consideration of the expected $\sim$1\% flux uncertainty, small background, xxx\% energy threshold uncertainty, and xxx\% energy resolution.

\begin{table}[t]
\label{deep_params}
  \begin{center}
    {\footnotesize
      \begin{tabular}{|c|c|}
      \hline
      \multicolumn{2}{|c|}{GEODM Assumptions} \\
      \hline 
        Scenarios considered& ``Baseline" and ``Optimistic"   \\
        $\nu$ source& $4\times10^{22}$ $\nu$/flavor/year w/ 13\% duty factor \\ 
        $\nu$ flux uncertainty & 2\%   \\ 
        Distance from $\nu$ source&2.3~km\\ 
        Exposure&4.5~ton$\cdot$year \\ \hline

      \end{tabular} 
      \caption{The assumptions used in the text for coherent neutrino detection with GEODM deep underground.}
}
\end{center}

\end{table}

\subsection{The effect of the coherent background on WIMP sensitivity}
\label{sec:WIMPsearch}

\begin{figure}[htp!]\begin{center}
%{\includegraphics[width=3.5in]{LimitPlotBaseline.pdf}}
\subfigure[ Cross section limit assuming ``baseline" GEODM efficiency.]{\label{Fig:WIMPLimitsBaseline}\includegraphics[width=3.4in]{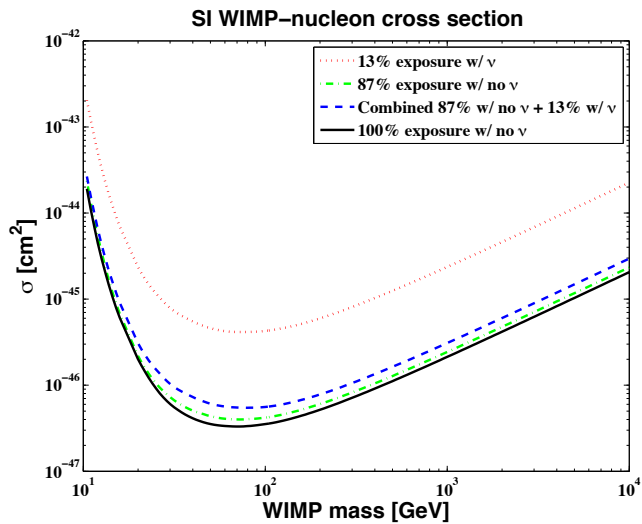}}
\subfigure[ Cross section limit assuming ``optimistic" efficiency.]{\label{Fig:WIMPLimitsOptimistic}\includegraphics[width=3.4in]{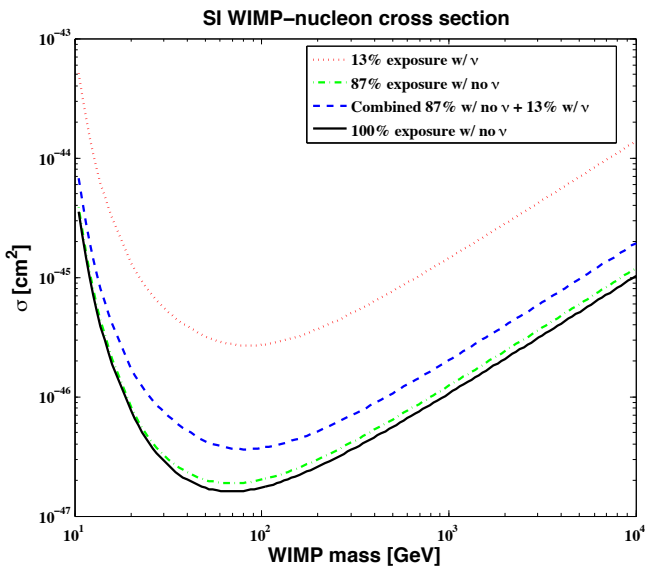}}
\end{center}
%\vspace{-3cm}
\caption{Expected average limits for the WIMP-nucleon cross section for a 4.5~ton$\cdot$year exposure, assuming no WIMP signal and calculated using the optimum interval method.  All limits include neutron and surface event leakage background events.  The `13\% exposure w/ $\nu$' limit also includes background events from coherent neutrino scattering, while `13\% exposure w/ no $\nu$', `87\% exposure w/ no $\nu$', and `100\% exposure w/ no $\nu$' do not.  The `Combined 87\% w/ no $\nu$ + 13\% w/ $\nu$' limit is obtained by combining the events and exposures from the `13\% exposure w/ $\nu$' limit and the `87\% exposure w/no $\nu$' limit, and treating it as a single experiment.  Each limit is calculated 100 times with background events randomly drawn from their distributions.  The resulting limits are averaged and averages are shown in this figure.}
\end{figure}

Since coherent neutrino scattering is an irreducible background for WIMP searches, the presence of a neutrino source near a dark matter experiment will reduce the sensitivity of the experiment.  In calculating a WIMP-nucleon cross section limit, one can either use data from only the 87\% of the time when the neutrino source is off or data from the 87\% beam-off time and the 13\% beam-on time.  Since background events reduce the sensitivity of a limit in the optimum interval method~\cite{OptimumInterval}, using the combined exposure may result in a worse limit than using only the beam-off exposure.  Fig. \ref{Fig:WIMPLimitsBaseline} and Fig. \ref{Fig:WIMPLimitsOptimistic} present a comparison of limits for the baseline and optimistic GEODM detector scenarios.  The limits assume a GEODM-style detector with a raw exposure of 4.5 ton$\cdot$year before efficiency cuts and the same assumptions as in the previous section.  In particular, we use the hypothetical efficiency curves shown in Fig.~\ref{efficiency}, a neutron rate of $0.15$~events/ton/year before efficiency convolution, and a total surface-event leakage of 1 event per 4.5 ton$\cdot$year raw exposure.  We use the recoil energy distribution from Fig.~\ref{efficiency} for neutrino events in the 13\% beam-on time and a detector at 2.3 km from a DAR neutrino source.

Using these parameters, we randomly generate 100 realizations of the background events and compute the average of the upper limits on the WIMP-nucleon cross section for each.  In this study, the beam-off data actually has greater sensitivity than the combined beam-on and beam-off data.  Using the beam-off data with 87\% exposure only, the mean upper limit at maximum sensitivity is $4.0 \cdot 10^{-47}$cm$^2$ (baseline) or $1.9 \cdot 10^{-47}$cm$^2$ (optimistic).  The combined data give a limit of $5.5 \cdot 10^{-47}$cm$^2$ (baseline) or $3.6 \cdot 10^{-47}$cm$^2$ (optimistic).  The results for the 87\% exposure obviously also have worse sensitivity than 100\% exposure without any beam-on time.  If there were no beam-on time, the upper limit at maximal sensitivity would be $3.3 \cdot 10^{-47}$cm$^2$ (baseline) or $1.7 \cdot 10^{-47}$cm$^2$ (optimistic).

Note that when the number of background events is zero, the sensitivity to WIMPs scales as the reciprocal of the exposure.  This occurs because the expected number of events in an experiment is proportional to the product of the cross section $\sigma$ and the exposure $E$,
\begin{equation}\label{eqn:CrossSectionScaling}
\mu \propto \sigma E.
\end{equation}
A limit at confidence level $C$ is obtained by determining the expected number of events $\mu$ such that there is a probability $C$ of observing zero events.  Assuming Poisson statistics the expected number of events is $\mu = -\log (1-C)$.  It  is thus apparent from Eq.~(\ref{eqn:CrossSectionScaling}) that the cross section limit $\sigma$ corresponding to confidence level $C$, scales as the reciprocal of the exposure $E$.  In the presence of a nonzero background that is proportional to exposure, the limit will scale more slowly than the reciprocal of the exposure.  Fixed backgrounds set by fiducial cuts may produce more complicated scaling behavior.

\section{Physics with a Detector Close to the Neutrino Source}
\label{sec:closedetector2}
A suitable detector within tens of meters of the stopped pion source could gather a rather large sample of elastic neutrino coherent scatters for physics studies. The 300~ft adit at DUSEL could provide direct tunnel access to an envisioned experimental site for such a detector. The cyclotron would then be located just outside the tunnel in a building against the cliff face. As discussed earlier, a coherent cross section measurement is sensitive to a number of physics possibilities. For simplicity, we consider a flux-integrated total cross section measurement as our figure of merit. However, the shape of the cross section as a function of energy is also interesting, especially in the case that a measured total cross section is inconsistent with expectation.

Although many low-threshold nuclear-recoil-sensitive detector technologies
could work (including a noble liquid detector~\cite{CLEAR,CLEAR2} or other dark matter detector technology),  for concreteness we consider the specific example of a set of GEODM-derived detectors with 16.7~kg raw mass, 20~m away from the stopped-pion neutrino source. With the optimistic efficiency estimate (Figure~\ref{efficiency}) and a fiducial mass of 10~kg, we expect a detected coherent rate of 0.74~events/(10~kg$\cdot$day) within the 13\% beam-on time window. The background rate design goal for such an experiment would be $<$0.1~events/(10~kg$\cdot$day) within the same window. This rate seems reasonable with $\sim$300~ft of rock shielding along with modest passive/active shielding immediately surrounding the detector for prompt cosmic-ray-induced background attenuation/tagging. The radiogenic background is assumed negligible, at a rate consistent with GEODM deep underground. The uncertainty on the non-beam related background estimate will easily be statistics-dominated with an $in-situ$ background measurement during beam-off.  We also assume that there are no background events from neutrons produced by the DAR neutrino source at a 20~m baseline with 17~m rock and 3~m Fe shielding.  This assumption was justified by performing a Geant4~\cite{geant} simulation of an isotropic neutron source along a 20~m baseline, fitting the flux at various distances from the neutron source to the functional form
\begin{equation*}
F(z) = \frac{A e^{-z/\lambda}}{z^2},
\end{equation*}
where $A$ and $\lambda$ are fit parameters and $z$ is the distance along the baseline.  The rate from this fit was extrapolated to 20~m distance and a 50~kg$\cdot$year exposure, and the number of neutron events was found to be negligible.  The neutron flux and spectrum used were taken from the SNS source, which is similar to the DAR source discussed here \cite{NuSNSproposal}.

\begin{table}[t]
\label{close_params}
  \begin{center}
    {\footnotesize
      \begin{tabular}{|c|c|} 
      \hline   
       \multicolumn{2}{|c|}{GEODM Module Close to the $\nu$ Source Assumptions} \\
       \hline 
        Scenario considered& ``Optimistic"   \\ 
        Source& $4\times10^{22}$ $\nu$/flavor/year w/ 13\% duty factor \\ 
        $\nu$ flux uncertainty & 2\%   \\ 
        Distance from $\nu$ source&20~m\\ 
        Exposure&50~kg$\cdot$year  \\ 
        Background rate & 0.1 events/(10~kg$\cdot$day) in beam window   \\\hline

      \end{tabular} 
      \caption{The assumptions used in the text for coherent neutrino detection with a GEODM module close to the $\nu$ source.}
}
\end{center}

\end{table}

A 50~kg$\cdot$year exposure with the previously described experimental design would yield about 1350 coherent events, assuming a coherent cross section consistent with the Standard Model. With an optimistic 1\% uncertainty on the target mass, 0.1 background events/(10~kg$\cdot$day) with statistical-only error, 2\% absolute flux normalization uncertainty, and a 0.5\% uncertainty on the energy resolution near threshold, a flux-averaged total cross section measurement with $<$5\% (statistical and systematic) uncertainty would be achieved. The assumptions that went into the event rate and cross section measurement uncertainty estimates are summarized in Table~IV.

\section{Conclusions}
\label{sec:conclusion}
Coherent elastic neutrino-nucleus scattering has never been observed. Relevant for supernova evolution, supernova-burst neutrino detection, probing non-standard neutrino interactions, and measuring $\sin^{2}\theta_{W}$ with neutrinos at low-$Q$, among other topics~\footnote{Coherent scattering has even been envisioned as a nuclear reactor monitoring tool~\cite{reactor1}\cite{reactor2}.}, the process is very well predicted by the Standard Model and confirmation of the $\sim$5\% precision theoretical cross section prediction is needed.

Dark matter detectors can double as coherent neutrino scattering experiments as the products of WIMP and coherent scattering interactions are predicted to be nearly identical. In the case of a decay-at-rest neutrino source and a suite of dark matter experiments at the same site, the deep underground detectors there would merely need to receive a beam timing signal in order to participate in the coherent search. Furthermore, these detectors would receive a free dark matter detection consistency check in the form of non-WIMP rare events in a well-known time window--all with a modest cost to the WIMP analysis/exposure. The power of this consistency check is strongly dependent on the number of coherent neutrino events collected, a value which is expected to be fairly low in most configurations. In both optimistic and baseline detection scenarios, the best limit on the WIMP-nucleon cross section uses only data from the period when the DAR source is off. In the optimistic scenario, the cross section limit is only about 12\% weaker than if no neutrino source were present.  

A coherent neutrino interaction discovery in GEODM could be achieved with a 2~ton$\cdot$year exposure. About 2.0 detected coherent neutrino events/ton/year over a background of 0.03~events/ton/year are expected in a GEODM-style detector at a 2.3~km baseline, given optimistic assumptions for energy threshold and detection efficiency. Even in a conservative (baseline) scenario, with energy threshold and efficiency reasonably consistent with CDMS~II, evidence for coherent neutrino scattering could be obtained with a 4.5~ton$\cdot$year exposure. In addition, a 10~kg fiducial mass GEODM-derived detector brought within 20~m of the neutrino source could collect about 1350 events with a 50~kg$\cdot$year exposure. Such a sample would be good for a $<$5\% flux-averaged total cross section measurement uncertainty and significant tests of the Standard Model.

\begin{center}
{ {\bf Acknowledgments}}
\end{center}

The authors thank Jocelyn Monroe for discussions, and Chuck Horowitz
 for providing the form factors for use in our calculations. J.~S. thanks Bonnie Fleming for support. We thank the National Science
Foundation for support.

\end{document}